\documentclass[aoas,MSNbibl,nameyear,rotating,dvips]{arximspdf}
\usepackage{url,breakurl}
\doi{10.1214/14-AOAS722} 
\volume{8}
\issue{3}
\pubyear{2014}
\firstpage{1443}
\lastpage{1468}


\begin{document}
\begin{frontmatter}

\title{Bayesian sparse graphical models for classification with
application to protein expression data}
\runtitle{Bayesian sparse graphical models}

\begin{aug}
\author[A]{\fnms{Veerabhadran} \snm{Baladandayuthapani}\corref{}\ead[label=e1]{veera@mdanderson.org}\thanksref{T1,T3,m1}},
\author[A]{\fnms{Rajesh} \snm{Talluri}\ead[label=e2]{rtalluri@mdanderson.org}\thanksref{T3,m1}},
\author[B]{\fnms{Yuan} \snm{Ji}\ead[label=e3]{jiyuan@uchicago.edu}\thanksref{T2,m2}},
\author[C]{\fnms{Kevin R.} \snm{Coombes}\ead[label=e4]{kcoombes@mdanderson.org}\thanksref{m3}},
\author[D]{\fnms{Yiling} \snm{Lu}\ead[label=e5]{yilinglu@mdanderson.org}\thanksref{m1}},
\author[E]{\fnms{Bryan T.} \snm{Hennessy}\ead[label=e6]{bryanhennessy74@gmail.com}\thanksref{m4,T4}},
\author[F]{\fnms{Michael A.} \snm{Davies}\ead[label=e7]{madavies@mdanderson.org}\thanksref{m1}}
\and
\author[G]{\fnms{Bani K.} \snm{Mallick}\ead[label=e8]{bmallick@stat.tamu.edu}\thanksref{m5}}\vspace*{6pt}
\runauthor{V. Baladandayuthapani et al.}
\affiliation{The University of Texas M.D. Anderson Cancer Center\thanksmark{m1},
NorthShore University HealthSystem and University of Chicago\thanksmark{m2},\break 
The Ohio State University\thanksmark{m3},
Beaumont Hospital\thanksmark{m4} and\break 
Texas A\&M University\thanksmark{m5}}\vspace*{6pt}
\address[A]{V. Baladandayuthapani\\
R. Talluri\\
Department of Biostatistics\\
The University of Texas\\
\quad M.D. Anderson Cancer Center\\
Houston, Texas 77030\\
USA\\
\printead{e1}\\
\phantom{E-mail:\ }\printead*{e2}}
\address[B]{Y. Ji\\
NorthShore University HealthSystem\\
1001 University Place\\
Evanston, Illinois 60201\\
USA\\
\printead{e3}}
\address[C]{K. R. Coombes\\
Department of Biomedical Informatics\\
The Ohio State University\\
\quad Wexner Medical Center\\
Columbus, Ohio 77030\\
USA\\
\printead{e4}}
\address[D]{Y. Lu\\
Department of Systems Biology\\
The University of Texas\\
\quad M.D. Anderson Cancer Center\hspace*{24pt}\\
Houston, Texas 77030\\
USA\\
\printead{e5}\\}
\address[E]{B. T. Hennessy\\
Beaumont Hospital\\
Dublin\\
Ireland\\
\printead{e6}}
\address[F]{M. A. Davies\\
Department of Melanoma\\
\quad Medical Oncology\\
The University of Texas\\
\quad M.D. Anderson Cancer Center\hspace*{24pt}\\
Houston, Texas 77030\\
USA\\
\printead{e7}}
\address[G]{B. K. Mallick\\
Department of Statistics\\
Texas A\&M University\\
College Station, Texas 77843\\
USA\\
\printead{e8}}
\end{aug}
\thankstext{T1}{Supported in part by NIH Grant R01 CA160736 and the
Cancer Center Support Grant (CCSG) (P30 CA016672).}
\thankstext{T2}{Supported by NIH R01 CA132897.}
\thankstext{T3}{Equal contributors.}
\thankstext{T4}{Supported by TRA (translational research award-TRA-2010-8)
from the Health Research Board Ireland (HRB) and Science Foundation Ireland (SFI).}

\received{\smonth{2} \syear{2013}}
\revised{\smonth{10} \syear{2013}}

%
\begin{abstract}
Reverse-phase protein array (RPPA) analysis is a powerful, relatively
new platform that allows for high-throughput,
quantitative analysis of protein networks. One of the challenges that
currently limit the potential of this technology is the lack of methods
that allow for accurate data modeling and identification of related
networks and samples. Such models may improve the accuracy of
biological sample classification based on patterns of protein network
activation and provide insight into the distinct biological
relationships underlying different types of cancer. Motivated by RPPA
data, we propose a Bayesian sparse graphical modeling approach that
uses selection priors on the conditional relationships in the presence
of class information. The novelty of our Bayesian model lies in the
ability to draw information from the network data as well as from the
associated categorical outcome in a unified hierarchical model for
classification. In addition, our method allows for intuitive
integration of {a priori} network information directly in the model
and allows for posterior inference on the network topologies both
within and between classes. Applying our methodology to an RPPA data
set generated from panels of human breast cancer and ovarian cancer
cell lines, we demonstrate that the model is able to distinguish the
different cancer cell types more accurately than several existing
models and to identify differential regulation of components of a
critical signaling network (the PI3K-AKT pathway) between these two
types of cancer. This approach represents a powerful new tool that can
be used to improve our understanding of protein networks in cancer.
\end{abstract}

%
\end{frontmatter}

\mbox{}

\section{Introduction}\label{sec1}

\subsection{Protein signaling pathways in cancer}\label{sec1.1}
The treatment of cancer is rapidly evolving due to an improved
understanding of the signaling pathways that are activated in tumors.
Global profiling of DNA mutations, chromosomal copy number changes, DNA
methylations and gene expression have greatly improved our appreciation
of the heterogeneity of cancer [\citet
{nishi03,blower07,gaur07,shanka07,ehrich08}]. However, the
characterization of protein signaling networks has proven to be much
more challenging. Several reasons underscore the critical importance of
overcoming this challenge: first, changes in cellular DNA and RNA both
ultimately result in changes in protein expression and/or function,
thus, protein networks represent the summation of changes that happen
at the DNA and RNA levels. Second, research has demonstrated that many
of the most common oncogenic genetic changes activate proteins in
kinase signaling pathways.
Numerous studies of protein networks and expression analysis have shown
promising results. Due to the hyperactivation of kinase signaling
pathways, numerous kinase inhibitors have been used in clinical trials,
frequently with dramatic clinical activity. Inhibitors that target
protein signaling pathways have been approved by the U.S. Food and Drug
Administration for a variety of cancer types, including chronic
myelogenous leukemia, breast cancer, colon cancer, renal cell carcinoma
and gastrointestinal stromal tumors [as reviewed in \citet{davies06}].



Protein networks need to be assessed directly, as DNA or RNA analyses
often do not accurately reflect or predict the activation status of
protein networks. Many proteins are regulated by post-translational
modifications, such as phosphorylation or cleavage events, that are not
detected by the analysis of DNA or RNA. Several studies have also
demonstrated marked discordance between mRNA and protein expression
levels, particularly for genes in kinase signaling and cell cycle
regulation pathways [\citet{varam05,shanka07}]. It has been
demonstrated recently, in both cancer cell lines and tumors, that
different genetic mutations in the same signaling pathway can result in
significant differences in the quantitative activation levels of
downstream pathway effectors [\citet
{stemke08,davies09,vasudevan09,park10}]. Although these observations
support the suggestion that direct measurements are essential to
measure protein network activation, a number of studies have
demonstrated that signaling pathways are frequently regulated by
complex feed-forward and feedback regulatory loops, as well as
cross-talk between different pathways [\citet
{mirz09,zhang09,halaban10}]. Thus, developing an accurate understanding
of the regulation of protein signaling networks will be optimized by
approaches that: (1) assess multiple pathways simultaneously for
different tumor types and/or conditions, and (2) allow for the use of
rigorous statistical approaches to identify differential functional networks.

\subsection{Reverse-phase protein lysate arrays}\label{sec1.2}

As explained, there is a strong rationale for methods that will
directly assess the activation status of protein \mbox{signaling} networks in
cancer. Traditional protein assays include immunohistochemistry (IHC),
Western blotting, enzyme-linked immunosorbent assay (ELISA) and mass
spectroscopy. Although IHC is a very powerful technique for the
detection of protein expression and location, it is critically limited
in network analyses by its non- to semi-quantitative nature. Western
blotting can also provide important \mbox{information}, but due to its
requirement for relatively large amounts of protein, it is difficult to
use when comprehensively assessing protein networks, and also is
semi-quantitative in nature. The ELISA method provides quantitative
analysis, but is similarly limited by requirements of relatively high
amounts of specimen and by the high cost of analyzing large pools of
specimens. Mass spectroscopy is a powerful, quantitative approach, but
its utility is mainly limited by the cost and time required to analyze
individual samples, which limits the ability to run large sample sets
that are needed to appropriately assess characteristics of disease
heterogeneity and protein networks.
Reverse-phase protein array (RPPA) analysis is a relatively new
technology that allows for quantitative, high-throughput, time- and
cost-efficient analysis of {protein networks using small amounts of
biological} material [\citet{paweletz01}; \citet{tibes06}].

\begin{figure}

\includegraphics{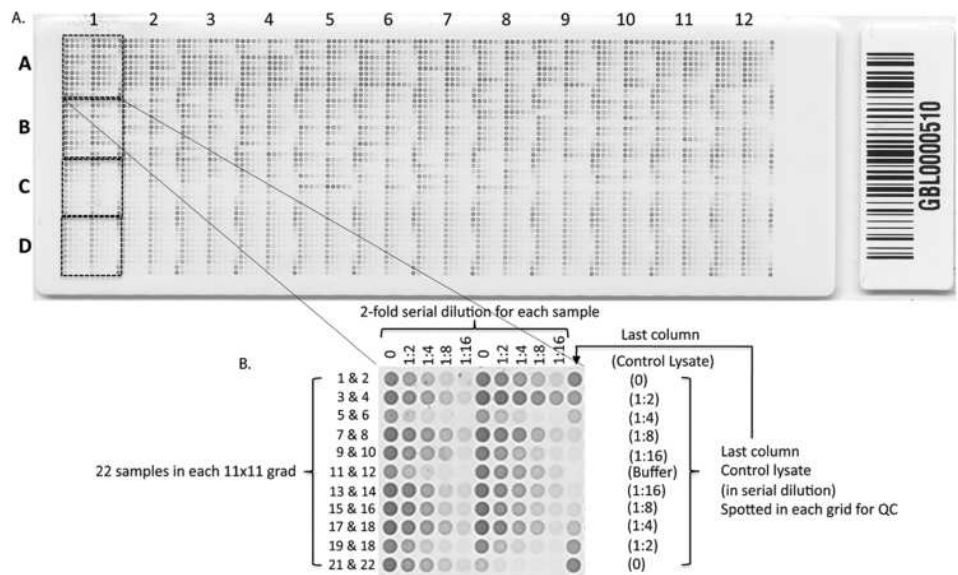}

\caption{An example of a reverse-phase protein array (RPPA) slide.
\textup{(A)}~Each slide is comprised of 4 rows (\textup{A--D}) of 12 columns (1--12) grids of 11X11 spots.
\textup{(B)}~Each grid has 22 individual samples and 11 controls.
Each row of the grid consists of~2 individual samples (each with 5
serial 2-fold dilutions) and one control spot. Reproduced with
permission from \protect\citet{tabchy2011}.}\label{figrppaslide}
\end{figure}

\subsubsection*{RPPA data collection}
{We provide a brief overview of the
RPPA experiment and data collection.} In order to perform RPPA,
proteins are isolated from the biological specimens such as cell lines,
tumors or serum using standard laboratory-based methods. The protein
concentrations are then determined for the samples and, subsequently,
serial 2-fold dilutions prepared from each sample are then arrayed on a
glass slide. Each slide is then probed with an antibody that recognizes
a specific protein epitope that reflects the activation status of the
protein. A~visible signal is then generated through the use of a signal
amplification system and staining. The signal reflects the relative
amount of that epitope in each spot on the slide, as shown in Figure~\ref{figrppaslide}. The arrays are then scanned and the resulting
images are analyzed with an imaging software specifically designed for
the quantification of RPPA analysis (MicroVigene, VigeneTech Inc.,
Carlisle, MA). The relative signal intensities of the dilution series
for each sample on the array are used to calculate the relative
{protein concentrations [\citet{neeley09,zhang09}].} Background
correction is used to separate the signal from the noise by subtracting
the extracted background intensity from the foreground intensity for
each individual spot. Relative protein amount is calculated using a
joint estimation method that utilizes {the logistic model of \citet
{tabus06}.} This method overcomes quenching at high levels and
background noise at low levels. An R package, SuperCurve, developed to
use with this joint estimation method is available at \url{http://bioinformatics.mdanderson.org/Software/OOMPA}. As with most
high-throughput technologies, the normalization of the resulting
intensities is conducted before any downstream analysis in order to
adjust for sources of systematic variation not attributable to
biological variation. Technical differences in protein loading for each
sample are determined by first dividing the results for each protein
measured by the average value among all the specimens, and then by
determining the average value for each sample across all of the
measured proteins. This relative loading factor is then used to
normalize the raw data for each sample, to correct for any differences
in protein loading between specimens. We refer the reader to \citet
{paweletz01} and \citet{hennessy2010} for more biological and
technical details concerning RPPAs.

%



Biological researchers typically choose specific targeted pathways
containing 50--200 proteins, usually assayed using the same number of
arrays, with each array hybridized against one protein. Because of the
reverse design (as compared to conventional gene expression
microarrays), RPPAs allow much larger sample sizes than the traditional
microarrays, thus allowing \textit{detailed} and \textit{integrated} analyses of protein signaling networks with higher
statistical power. Furthermore, this makes it possible to use RPPAs to
measure protein expression for multiple tumor classes and/or cell
conditions. The scientific aims we address using RPPA data in this
paper are threefold: to infer differential networks between tumor
classes/subtypes; to utilize {a priori} information in inferring
protein network topology within tumor classes/subtypes; and, finally,
to utilize network information in designing optimal classifiers for
tumor classification. We believe this will improve our understanding of
the regulation of protein signaling networks in cancer. Understanding
the differences in protein networks between various cancer types and
subtypes may allow for improved therapeutic strategies for each
specific type of tumor. Such information may also be relevant when
determining the origin of a tumor, which is clinically important in
cases with indeterminate histologic analysis, particularly for patients
who have more than one type of cancer.



\subsection{Graphical models for network analysis}\label{sec1.3}

A convenient and coherent statistical representation of protein
networks is accorded by graphical models [\citet{lauritzen96}]. By
``protein network'' we mean any graph with proteins as nodes, where the
edges between proteins may code for various biological information. For
example, an edge between two proteins may represent the fact that their
products interact physically (protein--protein interaction network), the
presence of an interaction such as a synthetic-lethal or suppressor
interaction [\citet{ryan05}], or the fact that these proteins code
for enzymes that catalyze successive chemical reactions in a pathway
[\citet{philippe03}].


Our focus is on undirected graphical models and on Gaussian graphical
models (GGM) in particular [\citet{whittaker90}]. These models
provide representations of the conditional independence structure of
the multivariate distribution---to develop and infer protein networks.
In such models, the nodes represent the variables (proteins) and the
edges represent pairwise dependencies, with the edge set defining the
global conditional independence structure of the distribution. We
develop an adaptive modeling approach for the covariance structure of
high-dimensional distributions with a focus on sparse structures, which
are particularly relevant in our setting in which the number of
variables/proteins ($p$) can exceed the number of observations ($n$).

GGMs have been under intense methodological development over the past few years in both frequentist
[\citet{meinshausen06,chaudhuri07,yuan2007,friedman08,bickel08}]
and Bayesian settings [\citet{giudici1999,roverato02,carvalho09}].
{\citet{wong2003} proposed a reversible jump MCMC-based Bayesian
model for covariance selection. In high-dimensional settings,
\citet{dobra03} used regression analysis to find directed acyclic
graphs and converted them to undirected (sparse) graphs to explore the
underlying network structure, and \citet{rod2011} proposed a new
approach for sparse covariance estimation in heterogeneous samples.}
However, most of the approaches we have cited focused on inferring the
conditional independence structure of the graph and did not consider
classification, which is one of the foci of our article. \citet
{rapaport07} used spectral decomposition to detect the underlying
network structure and classify genetic data using support vector
machines (SVM). More recently, \citet{monni10} proposed a
graph-based regression approach incorporating pathway information as a
prior for classification procedures, however, their method does not
detect differential networks based on available data. \citet
{zhu2009} proposed network-based classification for microarray data
using support vector machines. This was extended to network-based
sparse Bayesian classifiers by \citet{suarez2011}, but these
approaches do not estimate the network and also do not take into
account the differences in network structure between the two classes.
Another recent method is that of \citet{Fan2013}, who propose a
two-stage approach wherein they first select features and then
subsequently use the retained features and Fisher's LDA for
classification using only one covariance matrix for both the classes.

In this article, we propose a constructive method for sparse graphical
models using selection priors on the conditional relationships in the
presence of class information. Our method has several advantages over
classical approaches. First, we incorporate (integrate) the uncertainty
of the parameters in deriving the optimal rule via Bayesian model
mixing. Second, our network model provides an adaptively regularized
estimate of the covariance matrix and hence is capable of handling $n <
p$ situations. More importantly, our model uses this information in
deriving the optimal classification boundary. The novelty of our
Bayesian model lies in the ability to draw information from the network
data from all the classes as well as from the associated categorical
outcomes in a unified hierarchical model for classification. Through
this process, it offers the advantages of sparse Bayesian modeling of
GGM, as well as the simplicity of a Bayesian classification model. In
addition, with available online databases containing tens of thousands
of reactions and interactions, there is a pressing need for methods
integrating {a priori} pathway knowledge in the proteomic data
analysis models. We integrate prior information directly in the model
in an intuitive way such that the presence of an edge can be specified
by providing the probability of an edge being present in the
correlation matrix. Our method is fully Bayesian and allows for
posterior inference on the network topologies both within and between
classes. After fitting
the Bayesian model, we obtain the posterior probabilities of the edge
inclusion, which leads to false discovery rate (FDR)-based calls on
significant edges.


The structure of our paper is as follows. In Section~\ref{sec2} we outline our
Bayesian graph-based model for classification of RPPA data. Section~\ref{sec3}
focuses on Bayesian FDR-based determination of significant networks.
Section~\ref{sec4} presents the results of our case study using an RPPA
experiment. We end with a discussion and conclusion in Section~\ref{sec5}. All
technical details and additional analysis results are presented in the
supplementary material [\citet{suppl}].

\section{Probability model}\label{sec2}
Our data construct for modeling is as follows. We observe a tuple:
$(Z_i,\mathbf{Y}_i), i= 1,\ldots,n$, where $Z_i$ is a categorical outcome
denoting the type or subtype of cancer (binary or multi category) and
$\mathbf{Y}_i = (Y_i^{(1)},\ldots,Y_i^{(p)})$ is a $p$-dimensional vector
of proteins assayed for the $i$th sample/patient/array. We detail the
model here for binary classification (when $Z_{i}$ is a binary
variable), noting that generalization to multi-class classification can
be achieved in an analogous manner.
We factorize the joint distribution (likelihood) of the data $p(\mathbf
{Y_{i}},Z_{i})$, $ \forall i$ in the following manner
\[
p(\mathbf{Y_{i}},Z_{i})=p(\mathbf{Y_{i}}|Z_{i})p(Z_{i}),
\]
where the first component models the joint distribution of the $p$
proteins given the class variable $Z_i$ and the second component models
the marginal distribution of the class variables. We model the first
component as a mixture of the multivariate normal distributions as
\[
p(\mathbf{Y_{i}}|Z_{i},\bolds\mu,\bolds\Omega) \sim Z_{i}
N\bigl(\bolds\mu^{(1)},{\bolds \Sigma}^{(1)}\bigr)+(1-Z_{i})N
\bigl(\bolds\mu^{(2)},{\bolds\Sigma}^{(2)}\bigr),
\]
where ${\bolds\mu}^{(\bullet)}$ and ${\bolds\Sigma}^{(\bullet)}$ are the
corresponding means and covariances for the two classes. To specify the
marginal component, we note that in the classification framework only a
fraction of $Z$'s, say $Z^{u}$, will be unobserved (specifically in the
case of prediction, as shown in Section~\ref{sec2.2}) and they will be further
modeled as
\[
p\bigl(Z^{u}|h\bigr)\sim\operatorname{Bernoulli}(h),
\]
where we assign a Beta prior on probability $h$ as $h \sim\operatorname
{Beta}(\eta,\zeta)$. Note that this prior can be generalized to be
class-specific by allowing $h$ to depend on the class $k$ by changing
the corresponding hyperparameters $\eta_k,\zeta_k$.

Our main constructs of interest in this framework are $({\bolds\mu
}^{(k)},{\bolds\Sigma}^{(k)})$, $ k=1,2$ for each of the classes, where the
latter provides a dependence structure between the proteins, which we
model in a GGM framework. The key idea behind GGMs is rather to model
the precision matrix $\bolds{\Omega^{(k)}} = \bolds{\Sigma}^{(k)^{-1}} $,
which dictates the network structure between the variables. In this
framework of particular interest is the identification of zero entries
in the precision matrix---a zero entry at the $ij$th element of
$\Omega$ indicates conditional independence between the two random
variables $\mathbf{Y}_{i}$~and~$\mathbf{Y}_{j}$, given all other variables.
This is often referred to as the covariance selection problem in GGMs
[\citet{dempster72,cox96}]. In the section below we provide a
constructive method for sparse estimation (identification of many
zeros) of the precision matrix in high-dimensional settings, but also
allow for borrowing strength between classes to estimate the class-specific precision matrices for conducting classification.

\subsection{Parameterization of the precision matrix}\label{sec2.1}

Given the number of variables~$p$, the size of the precision matrix
($p\times p$) is potentially of high dimension. Instead of specifying a
global (joint) distribution on the precision matrix, we explore local
dependencies by breaking it down into components. For some
applications, it is desirable to work directly with standard deviations
and correlations [\citet{barnard2000,liechty03}] that do not
correspond to any type of parameterization (e.g., Cholesky, etc.).
{This parameterization has a practical motivation because most
biologists think in terms of correlations between the proteins, thus
easing prior elicitation, as we show below.}
To this end, we parameterize the precision matrix (for each class $k$,
suppressing the superscript for ease of notation) as $\bolds\Omega= \mathbf
{S} \times\mathbf{C} \times\mathbf{S}$, where $\mathbf{S}$ is a diagonal matrix
with nonzero diagonal elements that contains the inverse of the partial
standard deviations and $\mathbf{C}$ is a matrix that contains partial
correlation coefficients. Note that the correlation matrix $ \mathbf{C}$
satisfies the properties of a correlation matrix, that is, the partial
correlation coefficients ($\rho_{ij}$) between variables $i,j$ share a
one-to-one correspondence to the elements $C_{ij}$ as
\[
\rho_{ij} = \frac{-\Omega_{ij}}{(\Omega_{ii}\Omega_{jj})^{{1}/{2}}} = -C_{ij}.
\]

Due to this correspondence, sparse estimation of $\bolds{\Omega}$ directly
implies the identification of zeros in the elements of $\mathbf{C}$. Thus,
we model
$\mathbf{C}$ as a convolution,
\[
\mathbf{C} = \mathbf{A}\odot\mathbf{R},
\]
where $\odot$ is the Hadamaard operator indicating element-wise
multiplication between the two (stochastic) matrices: a \textit{selection}
matrix $\mathbf{A}$ and the corresponding \textit{correlation} matrix $\mathbf
{R}$ with the following properties:
\begin{itemize}
\item Both $\mathbf{A}$ and $\mathbf{R}$ are symmetric.
\item Both $\mathbf{A}$ and $\mathbf{R}$ have ones as their diagonal elements.
\item The off-diagonal elements of $\mathbf{A}$ are either 0 or 1 and the
off-diagonal elements of $\mathbf{R}$ lie in the range $[-1,1]$.
\item Both $\mathbf{A}$ and $\mathbf{R}$ \textit{need not} be positive
definite, but the convolution $\mathbf{C}$ \textit{has to be} positive definite.
\end{itemize}
%

In essence, $\mathbf{A}$ is a binary selection matrix that selects which of
the elements in $\mathbf{R}$ are zero or nonzero. In other words, $\mathbf{A}$
performs variable selection on the elements of the matrix $ \mathbf{R}$ by
shrinking the nonrequired variables (edges) exactly to zero and thus
inducing sparsity in the estimation of the resulting precision matrix
governing the GGM. We discuss hereafter the estimation and prior
specifications for each of these matrices.


\subsubsection*{Prior construction} $\mathbf{R}$ is a matrix
with all of its off-diagonal elements in the range $[-1,1]$, therefore,
we assign an independent uniform prior over $[-1,1]$ for all
$R_{ij}$, $i<j$. Correspondingly, since the off-diagonal elements of $\mathbf
{A}$ are binary (0~or~1), we assign an independent Bernoulli prior with
probability $q_{ij}$ for the element $A_{ij}$, $i<j$. Note that this
element-wise prior specification on $\mathbf{A}$ and $\mathbf{R}$ does not
ensure that the $\mathbf{C}$ $({=} \mathbf{A}\odot\mathbf{R})$ is positive definite---hence a valid graph. Thus, a key ingredient of our modeling scheme is
that we need an additional constraint: $\mathbf{C}\in\mathbb{C}_{p}$ where
$ \mathbb{C}_{p}$ is the space of all proper correlation matrices of
dimension $p$, such that
the joint convolved prior on $\mathbf{A}$ and $\mathbf{R}$ can be written as
\[
\mathbf{A},\mathbf{R}|\mathbf{q} \sim\prod_{i<j} \bigl\{
\operatorname {Uniform}_{R_{ij}}[-1,1] \operatorname{Bernoulli}_{A_{ij}}(q_{ij})
\bigr\}I(\bolds{\mathbf{A}\odot\mathbf {R}}\in\mathbb{C}_{p}),
\]
where $I(\bullet)$, the indicator function, ensures that the
correlation matrix is positive definite and introduces dependence among
the elements of the matrices $\mathbf{R},\mathbf{A}$, and $q_{ij}$ is the
probability of the $ij$th element being selected as 1.

We ensure the positive-definiteness constraint in our posterior
sampling sche\-mes. Specifically, we perform MCMC sampling in such a way
that the constraint $\mathbf{C}\in\mathbb{C}_{p}$ is satisfied---to search
over the possible space of valid correlation matrices. To implement the
constraint, we draw $R_{ij}, A_{ij}$, sequentially conditioned on all
other elements of $\mathbf{R}$ and $\mathbf{A}$ such that the realized value of
$C_{ij}$ ensures $\mathbf{C}$ is positive definite given all other
parameter values. Briefly, we follow the method of \citet
{barnard2000} to find the range $[u_{ij},v_{ij}]$ on the individual
elements of $R$ that will guarantee the positive definiteness of $\mathbf
{C}$. The resulting form of the conditional prior on the off-diagonal
elements $R_{ij}$ can be written as
\[
R_{ij}|a_{ij},A_{-ij},R_{-ij} \sim \operatorname{Uniform}(u_{ij},v_{ij})I(-1<R_{ij}<1),\qquad i\neq j,
i<j,
\]
where $R_{-ij}$ contains all other off-diagonal elements of $\mathbf{R}$
except the $ij$th element and $A_{-ij}$ contains all elements of $\mathbf
{A}$ except the $ij$th element. The limits of the Uniform distribution
$u_{ij}$ and $v_{ij}$ are chosen such that $\mathbf{C} = \mathbf{A}\odot\mathbf
{R}$ is positive definite and (conditionally) $u_{ij}$ and $v_{ij}$ are
functions of $R_{-ij}$ and $A_{-ij}$ (see Appendix A in the
supplementary material [\citet{suppl}] for the detailed proof).


In this construction, the parameter probability $q_{ij}$ controls the
degree of sparsity in the GGM in an adaptive manner by element-wise
selection of the entries of the correlation matrix. We assign a beta
hyperprior for the probabilities $q_{ij}$ as
\[
q_{ij} \sim \operatorname{Beta}(a_{ij},b_{ij}),\qquad i
\neq j,
\]
where the hyperparameters $a_{ij},b_{ij}$ can be set to induce prior
information on the graph structure (see Section~\ref{sec2.3}). To complete the
hierarchical specification, we choose an (exchangeable) inverse-gamma
prior on the inverse of the partial standard deviations $S$, which is a
diagonal matrix containing entries $S_i={\Omega}_{ii}^{{1}/{2}}$ as
$S_i\sim IG(g,h)$, $i = 1,2,\ldots,p$.

\subsubsection*{Borrowing strength between classes} Note that
in the above construction all the parameters are class-specific, that
is, are different for each class $k$, and thus model fitting and
estimation can be done for each class separately.
But the main advantage of Bayesian methodology lies in borrowing
strength between the classes for both estimation of the graphical
structure and subsequent prediction/classification. This can be
accomplished by having a variable that introduces dependence between
the classes linking the selection matrix $\mathbf{A}$. We introduce a
latent variable $\lambda_{ij}$ defined as
\[
\lambda_{ij} = \cases{ 1, &\quad if ${A}_{ij}^{1}
\neq{A}_{ij}^{2}$,
\vspace*{3pt}\cr
0, &\quad if ${A}_{ij}^{1}
= {A}_{ij}^{2}$,}
\]
where $\mathbf{A}^{1}$ and $\mathbf{A}^{2}$ are the class-specific selection
matrices. The binary variables $\lambda_{ij}$'s imply the presence or
absence of the same edge in the graphical model of both classes. In
other words, $\lambda_{ij}=1$ signifies a \textit{differential} edge
(i.e., the relation between the covariates $ i,j$ is significant in
only one class but not the other), whereas $\lambda_{ij}=0$ signifies a
\textit{conserved} edge (i.e., the relation between the covariates $ i,j$
is significant in both classes). Thus, the $\lambda$'s serve a dual
purpose in our model setup. They not only introduce dependence between
the classes, since they are shared between both classes, but also have
a distinct interpretation in terms of differential/conserved patterns
of dependence between the graphs for the classes. This information is
vital for understanding the biological processes and inferring
conclusions from the analysis, as we show in Section~\ref{sec4}.

Since the $\lambda_{ij}$'s are binary random variables, we propose a
Bernoulli prior on $\lambda_{ij}$ as
\[
\lambda_{ij}\sim\operatorname{Bernoulli}(\pi_{ij}),\qquad i<j,
\]
where the parameter $\pi_{ij}$ is the probability that the relation
between the $i$th and $j$th variables is different. We further
assign a beta hyperprior for the probabilities $\pi_{ij}$ as
\[
\pi_{ij} \sim \operatorname{Beta}(e_{ij},f_{ij}),\qquad i\neq j.
\]

To complete the prior specification on the graphical model, we propose
a normal prior on the means $(\bolds\mu^{(1)},\bolds\mu^{(2)})$ as
\[
\bolds\mu^{(k)}\sim N \bigl(\bolds\mu_0^{(k)},{
\mathbf{B}_0^{-1}}^{(k)} \bigr),\qquad k=1,2.
\]

\subsection{Prediction}\label{sec2.2}

In this section we lay out our graph-based prediction (classification)
scheme. Suppose the class variables $\mathbf{Z}$ (of size $n \times1 $) are
partitioned into a vector of training samples $\mathbf{Z}^{t}$ (of size
$n_t \times1 $) and a vector of (unknown) test/validation cases $\mathbf
{Z}^{u}$ (of size $n_u \times1 $) to be predicted. The corresponding
observed variables are also partitioned as $[\mathbf{Y}^{t};\mathbf{Y}^{u}]$.
Denote the observed data by $\mathcal{D} = \{\mathbf{Y}^t,\mathbf{Z}^t,\mathbf
{Y}^u\}$. In Bayesian prediction, for a new sample with protein
expression information $\mathbf{Y}^{u}$, we have to obtain the posterior
predictive probability that its class variable $\mathbf{Z}^{u}$, given all
observed data $\mathcal{D}$, is $p(\mathbf{Z}^{u}|\mathcal{D})$.

To estimate these probabilities, we treat $\mathbf{Z}^{u}\equiv\{Z_o^u\dvtx
o=1,\ldots,n_u\}$ as a parameter in the model and extend the MCMC
analysis to sample these values at each iteration. Specifically, we
draw $\mathbf{Z}^{u}$ from the corresponding conditional posterior
distribution in each MCMC iteration (see Appendix B in the
supplementary material [\citet{suppl}] for the full conditional distribution). The way
our model is specified, the posterior distribution of $\mathbf{Z}^{u}$ is
analyzed conditional not only on all the data from both classes
$\mathcal{D}$, but also on the parameters
that are shared between the classes. Thus, the predictions are obtained
in a single MCMC fitting procedure along with all other parameters,
thereby accounting for all sources of variation. We note that the
limitation of this method is that training and test splits of the data
must be contemplated prior to analysis (as is usually done) and/or
analysis fully repeated if new predictions are required.


The complete hierarchical formulation of our graph-based binary
classification model can be succinctly summarized as shown hereafter.
In addition, the directed acyclic graph (Figure~6 in the supplementary
material [\citet{suppl}]) shows a graphical representation of our model where the
circles indicate parameters and the squares observed random variables:
\begin{eqnarray*}
\mathbf{Y} &=& \bigl[\mathbf{Y}^{t},\mathbf{Y}^{u}\bigr] \sim \mathbf{Z} N
\bigl(\bolds\mu^{(1)},{\bolds\Omega ^{-1}}^{(1)}\bigr)+(1-
\mathbf{Z})N\bigl(\bolds\mu^{(2)},{\bolds\Omega^{-1}}^{(2)}\bigr),
\\
\mathbf{Z} &=& \bigl[\mathbf{Z}^{t}, \mathbf{Z}^{u}\bigr],
\\
{Z}_o^{u}&\sim& \operatorname{Bernoulli}(h_o),
\\
h_o&\sim&\operatorname{Beta}(\eta,\zeta),
\\
\bolds\mu^{(k)}&\sim& N\bigl(\bolds\mu_0^{(k)},{
\mathbf{B}_0^{-1}}^{(k)}\bigr),
\\
\bolds{\Omega}^{(k)}&=&\mathbf{S}^{(k)}\bigl(\mathbf{A}^{(k)}
\odot\mathbf{R}^{(k)}\bigr)\mathbf {S}^{(k)},
\\
\mathbf{A}^{(k)},\bolds\lambda,\mathbf{R}^{(k)}|\mathbf{q}^{(k)},\bolds
\pi&\sim&\prod_{i<j}\operatorname{Uniform}(u_{ij},v_{ij})
\operatorname {Bernoulli}\bigl(q_{ij}^{(k)}\bigr)
\\
&&\hspace*{13pt}{}\times
\operatorname{Bernoulli}(\pi_{ij}) I\bigl(\mathbf {C}^{(k)}\in
\mathbb{C}_{p}\bigr),
\\
q_{ij}^{(k)}&\sim& \operatorname{Beta}\bigl(
\alpha_{ij}^{(k)},\beta _{ij}^{(k)}\bigr),
\\
\pi_{ij} &\sim& \operatorname{Beta}(e_{ij},f_{ij}),\qquad i\neq j,
\\
S_i^{(k)}&\sim&IG(g,h), 
\end{eqnarray*}
where $k = 1,2$ corresponds to the two classes, $i,j=1,\ldots,p$, and
$o=1,\ldots,n_u$ corresponds to the size of the test/validation sample.
The full conditional distributions for MCMC sampling of the model
parameters and random variables are provided in Appendix B in the
supplementary material [\citet{suppl}].


\subsection{Incorporating prior pathway information and hyperparameter settings}\label{sec2.3}

As we mentioned before, there exists a huge amount of literature (prior
knowledge) describing the functional behaviors of proteins, as
characterized in metabolic, signaling and other regulation pathways. We
formally incorporate this {a priori} knowledge in our model through
the hyperparameter settings on the prior specification of $q_{ij}$, the
probability that the edge between protein $(i,j)$ will be selected. In
particular, we impose an informative prior on $\pi(q_{ij}) \sim
\operatorname{Beta}(a_{ij},b_{ij})$ and set the hyperparameters
$a_{ij}$ and $b_{ij}$ such that the distribution has a higher mean to
reflect our prior knowledge of the presence of an edge. For example, one
could set the following:

\begin{itemize}
\item prior on $q_{ij}$ as $\operatorname{Beta}(2,10)$ with mean 0.17, if there is
biological evidence that the edge does not play an important role in
the pathway;
\item prior on $q_{ij}$ as $\operatorname{Beta}(10,2)$ with mean 0.83, if there is
biological evidence that the edge plays an important role in the pathway;

\item prior on $q_{ij}$ as $\operatorname{Beta}(2,2)$ with mean 0.5, if no prior
information is available.
\end{itemize}

The prior information incorporated in the $q_{ij}$'s from online
databases is assumed to represent normal conditions only. Information
on relations between proteins that is affected by an intervention
and/or mutation can be elicited by expert opinion (e.g., from a
biologist). Information on the edges of graphs that is perturbed by a
mutation can be incorporated formally through our prior on $\pi_{ij}$,
the probability that controls the differential/conserved edge between
two different conditions. We specify informative priors in a manner
analogous to that of $q_{ij}$ (as shown above) in cases where such
information exists by setting $e_{ij},f_{ij}$ similarly to $a_{ij}$ and
$b_{ij}$. Finally, for the hyperparameters of the variance components,
we obtain a vague inverse gamma prior by setting $(g,h)=1$ and set the
hyperparameters for the beta prior on $h_o$ to be diffuse using $(\eta,\zeta)=2$.
%

\section{FDR-based determination of significant networks}\label{sec3}
Once we apply the MCMC methods, we are left with posterior samples of
the model parameters that we can use to perform Bayesian inference. Our
objective is twofold: to detect the ``best'' network/pathway based on
the significance of the edges and also to detect differential networks
between treatment classes. Given $p$ proteins, our network consists of
$p(p-1)/2$ unique edges, which could be large even for a moderate
number of proteins. Therefore, we need a mechanism that will control
for these large-scale comparisons, discover edges that are significant
and also detect differential edges between classes. We accomplish this
in a statistically coherent manner using
false discovery rate (FDR)-based thresholding to find significant
networks and also to differentiate networks across samples.

The MCMC samples explore the distribution of possible network
configurations suggested by the data, with each configuration leading
to a different topology of the network based on the model parameters.
Some edges that are strongly supported by the data may appear in most
of the MCMC samples, whereas others with less evidence may appear less
often. There are different ways to summarize this information in the
samples. One could choose the most likely (posterior mode) network
configuration and conduct conditional inference on this particular
network topology. The benefit of this approach would be the yielding of
a single set of defined edges, but the drawback is that the most likely
configuration may still appear only in a very small proportion of MCMC
samples. Alternatively, one could use all of the MCMC samples and,
applying Bayesian model averaging (BMA) [\citet{hoeting99}], mix
the inference over the various configurations visited by the sampler.
This approach better accounts for the uncertainty in the data, leads to
estimators of the precision matrix with the smallest mean squared error
and should lead to better predictive performance in class predictions
[\citet{raftery97}]. We will use this Bayesian model averaging approach.

From our MCMC iterations, suppose we have $M$ posterior samples of the
corresponding parameter set $\{ A^{(m)}_{ij}, m=1,\ldots,M\}$, for
which the selection indicator of the $ij$th edge is in the model.
Suppose further that the model averaged set of posterior probabilities
is set $\mathcal{P}$, the ${ij}$th element\vspace*{1pt} of which $\mathcal{P}_{ij}=
M^{-1}\sum_m A^{(m)}_{ij}$ and is a $p\times p$-dimensional matrix.
Note that $1-\mathcal{P}_{ij}$ can be considered Bayesian \mbox{$q$-}values, or
estimates of the local
false discovery rate [\citet{storey03,newton04}], as they measure
the probability of a false positive if the $ij$th edge is called a
``discovery'' or is significant. Given a desired global FDR bound
$\alpha\in(0,1)$, we can determine a threshold $\phi_{\alpha}$ with
which to flag a set of edges $\mathcal{X}_\phi=\{(i,j)\dvtx \mathcal{P}_{ij}
\geq\phi_{\alpha}\}$ as significant edges.

The significance threshold $\phi_{\alpha}$ can be determined based on
classical Bayesian utility considerations such as those described in
\citet{muller04} and based on the elicited relative costs of
false-positive and false-negative errors or can be set to control the
average Bayesian FDR, as in \citet{morris08}. The latter is the
process we follow here. For example, suppose we are interested in
finding the value $\phi_\alpha$ that controls the overall average FDR
at some level $\alpha$, meaning that we expect that only $100\alpha\%$
of the edges that are declared significant are in fact false positives.
Let $\operatorname{vec}(\mathcal{P}) = [\mathcal{P}_{t}; t=1,\ldots,p(p-1)/2]$ be
the vectorized version of the set $\mathcal{P}$ containing the unique
posterior probabilities of the edges, stacked columnwise. We first sort
$\mathcal{P}_t$ in descending order to yield $\mathcal
{P}_{(t)},t=1,\ldots,p(p-1)/2$. Then $\phi_\alpha=\mathcal{P}_{(\xi)}$,
where $\xi=\max\{j^*\dvtx j^{*-1}\sum_{j=1}^{j^*} \mathcal{P}_{(t)} \le
\alpha\}$. The set of regions $\mathcal{X}_{\phi_\alpha}$ then can be
claimed to be significant edges based on an average Bayesian FDR of
$\alpha$.

This FDR-based thresholding procedure can be extended to find
differential networks between different populations (tumor
classes/subtypes), for example, to identify edges that are
significantly different between tumor types. To this end, we use the
corresponding parameter set $\{ \lambda^{(m)}_{ij}, m=1,\ldots,M\}$,
which is the selection indicator of the differential edge between the
$ij$th covariates in the model. The model averaged set of posterior
probabilities is set $\mathcal{P}^d$, the ${ij}$th element of which
$\mathcal{P}^d_{ij}= M^{-1}\sum_m \lambda^{(m)}_{ij}$. We use this same
procedure to arrive at a set of differential edges $\mathcal{X}_\phi=\{
(i,j)\dvtx \mathcal{P}^{d}_{ij} \geq\phi_{\alpha}\}$ with $\phi_{\alpha}$
chosen to control the Bayesian\vspace*{-2pt} FDR at level~$\alpha$. We use a similar
procedure on the parameter set $\{1-\lambda^{(m)}_{ij}, m=1,\ldots,M\}
$, to arrive at a set of common edges $\mathcal{X}_\phi=\{(i,j)\dvtx \mathcal
{P}^{c}_{ij} \geq\phi_{\alpha}\}$ with $\phi_{\alpha}$ chosen to
control the Bayesian FDR at level $\alpha$.

\section{Data analysis}\label{sec4}

\subsection{Classification of breast and ovarian cancer cell lines}\label{sec4.1}

Breast and ovarian cancer are two of the leading causes of
cancer-related deaths in women [\citet{jemal09}]. Both of these
diseases are frequently affected by mutations in kinase signaling
cascades, particularly those involving components of the PI3K-AKT
pathway [\citet{mills03,hennessy08,yuan08,bast09}]. The PI3K-AKT
pathway is one of the most important signaling networks in
carcinogenesis [\citet{vivanco02}] and is affected more than any
other signaling pathway by activating mutations in cancer tissues
[\citet{yuan08}]. Aggressive drug development efforts have
targeted this critical oncogenic pathway, and inhibitors of multiple
different components of the PI3K-AKT pathway have been developed and
are in various stages of preclinical and clinical testing [\citet
{hennessy05,courtney10}].

We applied our methodology to identify differences in the regulation of
the PI3K-AKT signaling network in breast and ovarian cancers. For this
analysis, we used expression data of $p=50$ protein markers in
signaling pathways from an RPPA analysis of human breast ($n_1=51$) and
ovarian ($n_2=31$) cancer cell lines grown under normal tissue culture
conditions [\citet{stemke08}]. We used the known connections in
the PI3K-AKT pathway
suggested by previous studies and expert opinion as {a priori}
information in our model, as stated in Section~\ref{sec2.3}.

%
\begin{sidewaysfigure}

\includegraphics{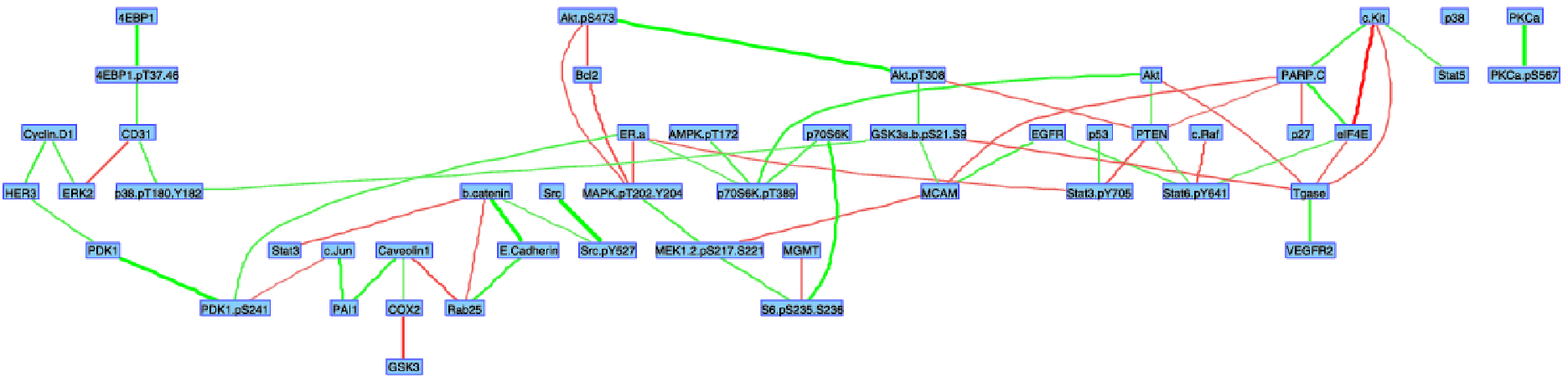}

\centering{\footnotesize{(a) Breast network}}\vspace*{6pt}

\includegraphics{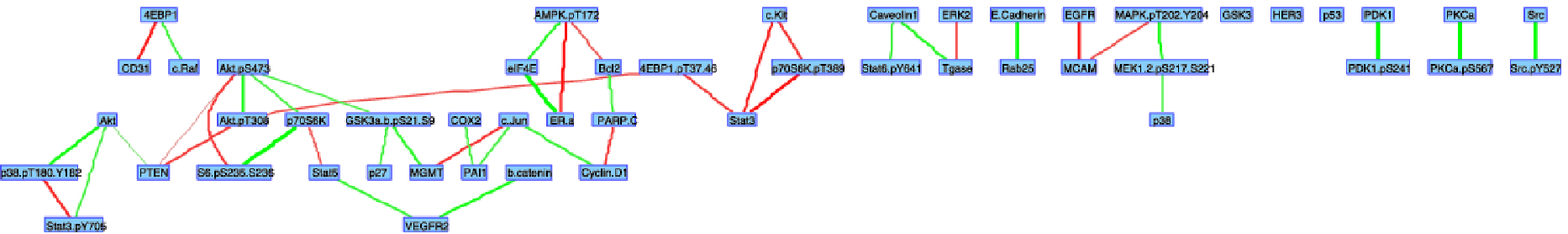}

\centering{\footnotesize{(b) Ovary network}}
\caption{Significant edges for the proteins in the PI3K-AKT kinase
pathway for breast \textup{(a)} and ovarian cancer cell lines \textup{(b)} computed using a Bayesian FDR of 0.10. The red (green) lines
between the proteins indicate a negative (positive) correlation between
the proteins. The thickness of the edges corresponds to the strength of
the associations, with stronger associations having greater thickness.}\label{brnet}
\end{sidewaysfigure}

\begin{figure}

\includegraphics{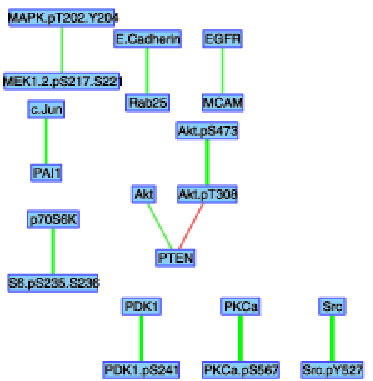}

\footnotesize{(a) Conserved network between ovarian and breast cancer cell lines}\vspace*{6pt}

\includegraphics{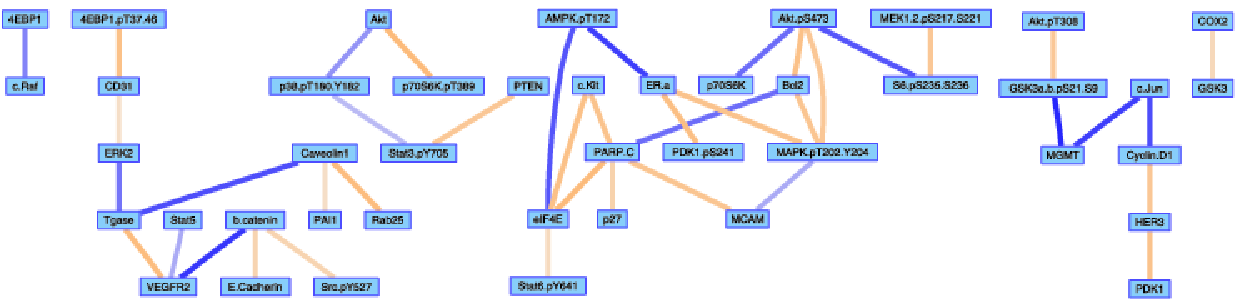}

\footnotesize{(b) Differential network between ovarian and breast cancer cell lines}
\caption{Conserved and differential networks for the proteins in the
PI3K-AKT kinase pathway between breast and ovarian cancer cell lines
computed using a Bayesian FDR set to 0.10. In the conserved network
(top panel), the red (green) lines between the proteins indicate a
negative (positive) correlation between the proteins.
In the differential network (bottom row), the blue lines between the
proteins indicate a relationship that was significant in the ovarian
cancer cell lines but not in the breast cancer cell lines; the orange
lines between the proteins indicate a relationship in the breast cancer
cell lines but not in the ovarian cancer cell lines. The thickness of
the edges corresponds to the strength of the associations, with
stronger associations having greater thickness.}\label{fig3}
\end{figure}

The significant networks based on a Bayesian FDR cutoff of $\alpha =
0.1$ for breast and ovarian cancer samples are shown in Figure~\ref{brnet}(a)~and~(b), respectively. The red edges indicate a
negative association (regulation) and the green edges indicate a
positive interaction between the proteins. The edges are represented by
lines of varying degrees of thickness based on the strength of the
association (correlation), with higher weights having thicker edges and
lower weights having thinner edges.
In order to identify biological similarities and differences between
the breast and ovarian cancer cell lines, we compared the results of
our network analyses of the two cancer types. Plotted in Figure~\ref{fig3}(a) are the conserved (common) edges between the two cancer
types. The differential network between the two cancer types,
controlling for a Bayesian FDR cutoff of $\alpha = 0.1$, is shown in
Figure~\ref{fig3}(b).

%

A number of protein--protein relationships demonstrated significant
similarity between the two cancer types. For example, both breast
cancer and ovarian cancer cell lines exhibited a marked negative
association between the \mbox{levels} of PTEN and phosphorylated AKT
(Akt.pT308). This relationship was expected due to the critical
regulation of 3-phopshatidylinositols by the lipid phosphatase activity
of PTEN, and has previously been demonstrated as a significant
interaction in multiple tumor types [\citeauthor{davies98} (\citeyear{davies98,davies99,davies09}),
\citet{stemke08,vasudevan09,park10}]. Although
this concordance was expected, our analysis also identified a large
network of \mbox{differential} protein interactions among the breast and
ovarian cancer cell lines [Figure~\ref{fig3}(b)]. In this figure,
the edges in blue indicate relationships between proteins that were
present in the ovarian cancer cell lines but not in the breast cancer
cell lines using our FDR cutoff, and the orange edges indicate
relationships present in the breast cancer cell lines but not in the
ovarian cancer cell lines. In addition, the thickness of the edges
corresponds to the strength of the association. Notable differential
connections in this analysis include the association of phosphorylated
AKT (Akt.pS473) with BCL-2 (Bcl2) and phosphorylated MAPK
(MAPK.pT202.Y204) in breast cancer. Both of these, BCL-2(Bcl2) and
phosphorylated (activated) MAPK (MAPK.pT202.Y204), may contribute to
tumor proliferation and survival, and are therapeutic targets with
available inhibitors. The \mbox{association} of different proteins with the
expression of the estrogen receptor, phosphorylated PDK1 (PDK1.pS241)
and MAPK (MAPK.pT202.Y204) in breast cancer and phosphorylated AMPK
(AMPK.pT172) in ovarian cancer, may also have therapeutic implications,
as the estrogen-receptor blockade is a treatment used in both advanced
breast and ovarian cancer.

We used this network information to build a classifier to distinguish
between breast cancer and ovarian cancer samples as explained in
Section~\ref{sec2}. We assessed the performance of the classifiers using
cross-validation techniques. {In particular, we generated 100 random
selections of training and test data sets with 66\% and 33\% splits of
the data, respectively. We fit our Bayesian graph-based classifier
(BGBC) and compared our method to six other methods: the network-based
support vector machine (SVM) [\citet{zhu2009}], $K$-nearest neighbor
(KNN), linear discriminant analysis (LDA), diagonal linear discriminant
analysis (DLDA), diagonal quadratic discriminant analysis (DQDA) and
naive Bayes classifier (NBC) [\citet{john1995}] methods. We
implemented the network-based SVM using the R package ``pathclass.'' The
network structure was specified to be the common network for the two
classes obtained from the BGBC algorithm, as this method does not
explicitly estimate the network. All other input parameters were set at
the default settings for the network-based SVM function. We implemented
all the other methods using the corresponding MATLAB functions.}

\begin{table}
\tabcolsep=0pt
\caption{Misclassification error rates for different classifiers for
ovarian and breast cancer data sets. The methods compared here are
SVM (network-based support vector machine),
LDA (linear discriminant analysis),
KNN ($K$-nearest neighbor),
DQDA (diagonal quadratic discriminant analysis),
DLDA (diagonal linear discriminant analysis),
NBC (naive Bayes classifier) and
BGBC (Bayesian graph-based classifier),
which is the method proposed in this paper with and without
incorporating prior information, denoted by BGBC (prior) and BGBC (w/o
prior), respectively. The mean and the standard deviation are values of
the misclassification percentage over 100 random splits of the data}\label{tabl3}
\begin{tabular*}{\tablewidth}{@{\extracolsep{\fill}}lcccccccc@{}}
\hline
& \textbf{SVM} & \textbf{KNN} & \textbf{LDA} & \textbf{DLDA} & \textbf{DQDA} & \textbf{NBC} & \textbf{BGBC} & \textbf{BGBC w/o prior} \\
\hline
Mean & 8.03 & 15.14 & 25.48 & 12.07 & 13.74 & 13.37 & 6.59 & 10.88  \\
SD   & 5.44 & \phantom{0}6.82 & 10.63& \phantom{0}5.829 & 6.70 & \phantom{0}6.96 & 4.06 & \phantom{0}6.31\\
\hline
\end{tabular*}
\end{table}

The average misclassification errors (along with standard errors)
across all splits for all the methods on the test set are shown in
Table~\ref{tabl3}. The BGBC method had much lower misclassification rates
compared to the other methods (the other methods ignore the underlying
network structure of the proteins). We believe that this improved
precision is due to the fact that the mean expression profiles of the
breast and ovarian cancer cell lines are very similar so there is not
enough information in the mean to classify the two cases. Hence,
means-based classifiers, especially KNN and LDA (both of which use
identity and diagonal covariances), underperform as compared to our
method. The results of the DQDA method could be a bit closer to those
of the BGBC method, but the former method ignores the
cross-connections, that is, network information, and hence results in a
higher misclassification rate. The QDA could not be performed because
the estimation of different covariance matrices for different classes
is an ill-posed problem for $n<p$. { We also tested the performance of
BGBC using prior information and without using prior information in
estimating the networks. The last two columns of Table~\ref{tabl3} show that
incorporating prior information improves our classification
performance. Furthermore, the inclusion of prior information leads to
sparser networks (as shown in Figure~7 in the supplementary material [\citet{suppl}]),
as the prior information provides information about important and
unimportant relationships, which aids our classification model. }

We further note that nonlinear (quadratic) boundaries are obtained by
using network information (since we model the covariance matrix),
whereas linear boundaries are obtained by ignoring the network
information (linear/diagonal \mbox{discriminant} based approaches). The
classification boundary (Figure~8 in the supplementary material [\citet{suppl}])
exemplifies our intuition and approach. We have a $p (=50)$-dimensional
quadratic classification boundary based on the GGM. In order to
visualize this, we projected the boundary and the data onto two
randomly selected dimensions/covariates. Two of those projections are
shown in the figure, which confirm our intuition of how nonlinear
boundary is more effective than a linear boundary in classification.

\subsection{Effects of tissue culture conditions on network topology}\label{sec4.2}

Cell lines derived from tumors are a powerful research tool, as they
allow for detailed characterization and functional testing. Genetic
studies support the concept that cell lines generally mirror the
changes that are detected in tumors, particularly at the DNA and RNA
levels [\citet{neve06}]. However, the activation status of
proteins can be impacted by the use of different environmental
conditions in the culturing of cells. A key scientific question in the
analysis of protein networks in cancer cell lines is the variability of
network topologies due to differing tissue culture conditions. In order
to assess if different network connectivity is observed under varying
culture conditions, we used three different tissue culture conditions
to grow the 31 ovarian cancer cell lines used in the previous analysis.

For condition ``A,'' the cells were grown in tissue culture media that
was supplemented with growth factors in the form of fetal calf serum
(5\% of the total volume), which is a standard condition for the
culturing of cancer cells. For condition ``B,'' the cells were harvested
after being cultured in the absence of growth factors (serum) for 24
hours. For condition ``C,'' cells were grown in the absence of growth
factors for 24 hours, then they were stimulated acutely (20 minutes)
with growth factors (5\% fetal calf serum). Proteins were harvested
from each cell line for each tissue culture condition. The experimental
procedure used for the isolation and RPPA analysis of proteins from the
cancer cells growing under normal, serum-replete tissue culture
conditions has been described previously [\citet{davies09,park10}]. Protein isolation, processing and RPPA analysis were
performed using the same methodology for all three conditions.

\begin{figure}

\includegraphics{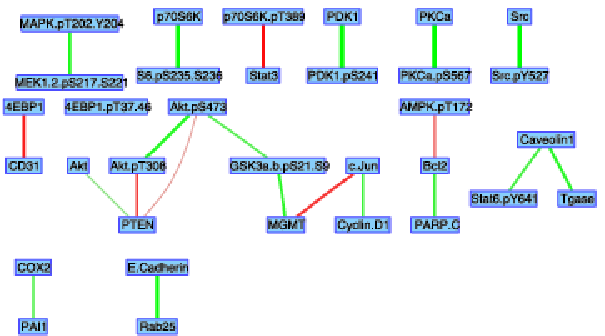}

\footnotesize{(a) Conserved network of Ovary A and Ovary B}\vspace*{6pt}

\includegraphics{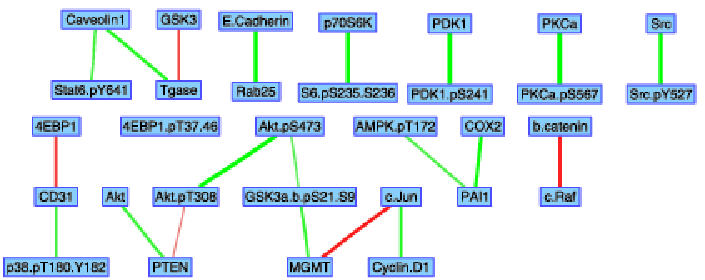}

\footnotesize{(b) Conserved network of Ovary B and Ovary C}\vspace*{6pt}

\includegraphics{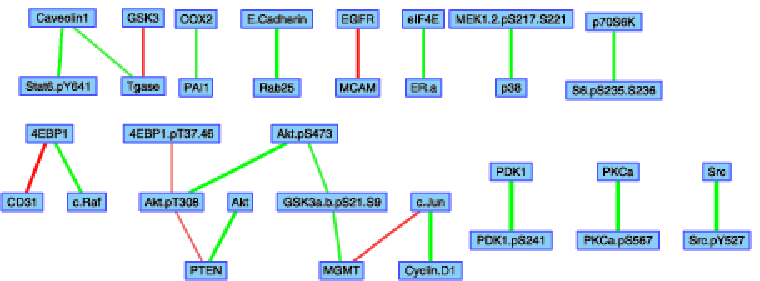}

\footnotesize{(c) Conserved network of Ovary A and Ovary C}\vspace*{6pt}

\begin{tabular}{@{}cccc@{}}

\includegraphics{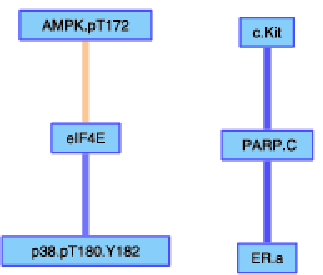}
 & \includegraphics{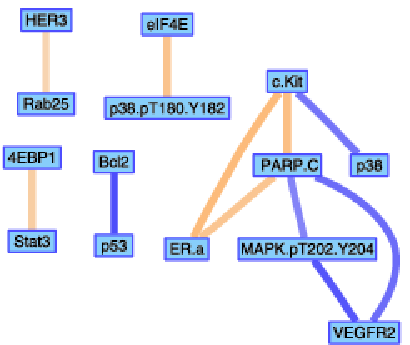} & \includegraphics{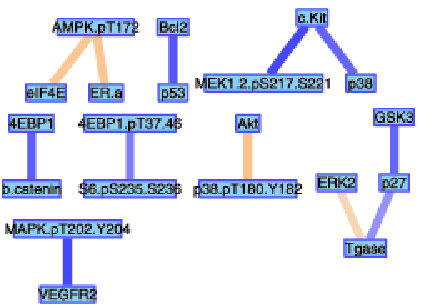}\\
\footnotesize{(d) Differential network}& \footnotesize{(e) Differential network} & \footnotesize{(f) Differential network}\\
\footnotesize{of Ovary A and Ovary B}& \footnotesize{of Ovary B and Ovary C} & \footnotesize{of Ovary A and Ovary C}
\end{tabular}
\caption{Conserved and differential networks for the proteins in the
PI3K-AKT kinase pathway between ovarian cancer cell lines grown in
three different tissue culture conditions: A, B and C (see main text)
computed using a Bayesian FDR set to 0.10. In the conserved network
[\textup{(a)--(c)}], the red (green) lines between the proteins indicate a
negative (positive) correlation between the proteins.
In the differential network [\textup{(d)--(f)}], the blue lines between the
proteins indicate a relationship that was significant in the ovarian
cancer cell lines but not in the breast cancer cell lines; the orange
lines between the proteins indicate a relationship in the breast cancer
cell lines but not in the ovarian cancer cell lines. The thickness of
the edges corresponds to the strength of the associations, with
stronger associations having greater thickness.}\label{fig4}
\end{figure}

The RPPA data for each condition were then analyzed for protein--protein
interactions using the GGM method. The topology maps for the ovarian
cancer cells for the A, B and C tissue culture conditions are shown in
Figure~12(a), (b) and~(c) (provided in the supplementary
material [\citet{suppl}]), respectively. We then performed comparisons of the results
based on each of the three conditions in order to identify protein
topology networks that were similar and different between each of the
tissue culture conditions. As conditions A (media replete with growth
factor) and B (media starved of growth factor) both represented
steady-state tissue culture conditions, we initially compared these
protein networks using a Bayesian FDR of 10\%. The networks that are
shared between the two conditions are shown in Figure~\ref{fig4}(a); the differential associations are presented in Figure~\ref{fig4}(d). We detected 21 significant protein interactions
that were common for conditions A and B, and 4 interactions that were
different. Thus, the overwhelming majority of protein--protein
associations that were observed were maintained regardless of the
presence or absence of growth factors (serum) in the tissue culture
media. We then compared the significant relationships identified for
condition B (media starved of growth factor) versus condition C (media
starved, then acutely stimulated with growth factor). This comparison
showed increased discordance of results, as we detected 20 associations
that were common for conditions B and C [Figure~\ref{fig4}(b)], but
11 associations that differed significantly [Figure~\ref{fig4}(e)]. Similarly, the comparison of networks between the A and
C conditions identified 22 shared protein interactions [Figure~\ref{fig4}(c)] and 12 differential interactions [Figure~\ref{fig4}(f)]. Of the differential interactions noted for the
comparisons of conditions B versus C and A versus C, only 2 were
observed in both comparisons (c-KIT and P38; VEGFR2 and
MAPK.pT202.Y204). Neither of these 2 relationships was among the
differential protein interactions in the analysis of condition A versus
condition B. Of the 4 relationships that differed in the comparison of
condition A versus condition B, 3 of the relationships were also
identified as differing significantly when comparing condition B versus
condition C (eIF4E and P38.pT180.Y182; c-Kit and PARP.cleaved;
PARP.cleaved and ER.alpha), and the fourth differed significantly for
the comparison of condition A versus condition C (AMPK.pT172 and eIF4E).
This analysis suggests that protein--protein relationships are largely
maintained under steady-state tissue culture conditions. However, these
interactions may differ significantly in the setting of acute growth
factor stimulation. We have included the explicit comparisons of our
inferred networks with the prior PI3K-AKT pathway in Figures~13--16 in
the supplementary material [\citet{suppl}]. The posterior means of the covariance
matrices corresponding to the networks are also now plotted as heat
maps in Figures~17--20 in the supplementary material [\citet{suppl}]. The exact
posterior mean estimates are also provided as excel files downloadable
from the corresponding authors' website at \url{http://odin.mdacc.tmc.edu/~vbaladan/Veera_Home_Page/Software_files/Covariance_Matrices.xlsx}.

%

\section{Discussion and conclusions}\label{sec5}
We present methodology to model sparse graphical models in the presence
of class variables in high-dimensional settings, with a particular
focus on protein signaling networks. Our methods allow for borrowing
strength between classes to assess differential and common networks
between the classes of cancer/tumor conditions. In addition, our method
allows for the
effective use of prior information about signaling pathways that is
already available to us from various sources to help in decoding the
complex protein networks. Improved understanding of the differential
networks can be crucial for biologists when designing their
experiments, allowing them to concentrate on the most important factors
that distinguish tumor types. Such information may also help to narrow
the drug targets for specific types of cancer.
Knowledge of the common networks can be used to develop a drug for two
different types of cancer that targets proteins that are active in both
types. Data on the differential edges may be used as a good screening
analysis, allowing researchers to eliminate unimportant proteins and
concentrate on effective proteins when designing advanced patient-based
translational experiments.

In this article we focused on undirected graphical models and not on
directed (casual) networks. Directed graphical models, such as Bayesian
networks and directed acyclic graphs (DAGs), have explicit causal
modeling goals that require further modeling assumptions. In our
formulation, we provide a natural and useful technical step in the
identification of high posterior probability undirected graphical
models, assuming a random sampling paradigm. In addition, our models
infer network topologies that assume a steady-state network. Some of
the protein networks may be dependent on causal relations between the
nodes, which would require us to model data over time to infer the
complete dynamics of the network.
We leave this task for future consideration.

With regard to computation time, our MCMC chains are fairly fast for
high-dimensional data sets such as those we considered, with a
5000-iteration run taking about 15 minutes. The source code, in MATLAB
(The Mathworks, Inc., Natick, MA), takes advantage of several matrix
optimizations available in that language environment. The
computationally-involved step is the imposition of a positive
definiteness on the correlation matrix. Optimizations to the code have
been made by porting some functions into C. The software is available
by contacting the first author.

Our main motivation for this work was to provide a constructive
framework to conduct classification using sparse graphical methods that
incorporate prior information. We assume parametric structures
(likelihood/priors) throughout for ease of interpretation and
computation, and our results indicate that this performs reasonably
well on both real and simulated data sets. Extending to nonparametric
settings would be an excellent avenue of future research that we would
wish to undertake.

\section*{Acknowledgments}
The content is solely the responsibility of the authors and does not
necessarily represent the official views of the National Cancer
Institute or the National Institutes of Health.

%


\begin{supplement}[id=suppA]
\stitle{Supplement to ``Bayesian sparse graphical models for classification with application to protein expression data''}
\slink[doi]{10.1214/14-AOAS722SUPP} 
\sdatatype{.pdf}
\sfilename{aoas722\_supp.pdf}
\sdescription{The supplementary material includes
Appendix~A: Positive definiteness constraint,
Appendix B: Full conditional distributions
and Appendix C: Simulations.}
\end{supplement}

%

\printaddresses
\end{document}